\documentclass[preprint,journal]{vgtc}            

\onlineid{0}
\preprinttext{}
\manuscriptnote{}

\vgtccategory{Research}

\title{Structured Visualization Design Knowledge for Grounding Generative Reasoning and Situated Feedback}

\author{%
  \authororcid{P\'{e}ter Ferenc Gyarmati}{0009-0006-6122-2709},
  \authororcid{Dominik Moritz}{0000-0002-3110-1053},
  \authororcid{Torsten M\"{o}ller}{0000-0003-1192-0710},
  and \authororcid{Laura Koesten}{0000-0003-4110-1759}
}

\authorfooter{
  \item P\'{e}ter Ferenc Gyarmati is with the University of Vienna, Austria, and Mohamed bin Zayed University of Artificial Intelligence, UAE.
  \item Dominik Moritz is with Carnegie Mellon University, USA.
  \item Torsten M\"{o}ller is with the Faculty of Computer Science, University of Vienna, Austria.
  \item Laura Koesten is with the the Mohamed bin Zayed University of Artificial Intelligence, UAE, the Faculty of Computer Science, University of Vienna, Austria, and the Austrian Institute of Technology, Austria.
}

\abstract{%
    Automated visualization design navigates a tension between symbolic systems and generative models. Constraint solvers enforce structural and perceptual validity, but the rules they require are difficult to author and too rigid to capture situated design knowledge. Large language models require no formal rules and can reason about contextual nuance, but they prioritize popular conventions over empirically grounded best practices. We address this tension by proposing a cataloging scheme that structures visualization design knowledge as natural-language guidelines with semantically typed metadata. This allows experts to author knowledge that machines can query. An expert study ($N=18$) indicates that practitioners routinely adapt heuristics to situational factors such as audience and communicative intent. To capture this reasoning, guideline sections specify not only the advice but also the contexts in which it applies, the exceptions that invalidate it, and the sources from which it derives. We demonstrate the scheme's expressiveness by cataloging 781 guidelines drawn from cognitive science, accessibility standards, data journalism, and research on rhetorical aspects of visual communication. We embed guideline sections in a vector space to enable semantic retrieval and structural analysis of knowledge. This reveals conflicting advice across sources and transferable principles between domains. We validate the scheme by evaluating whether catalog-grounding improves how generative models reason about what design to create, not just how to implement it. To support this, we enrich a natural language-to-visualization (NL2VIS) benchmark dataset with canonical queries that capture analytical intent while leaving the visual encoding open. A factorial evaluation across multiple frontier models, chart grammars, and contexts yields $1,260$ visualization instances from a stratified subset of this data. Results demonstrate that catalog-grounded generation improves design quality and reduces uncontrolled design variance. Rather than replacing constraint-based tools, our scheme provides what they lack: situated guidance that generative systems can retrieve to ground their reasoning, users can verify against cited sources, and experts can author as knowledge evolves.
}

\keywords{Human-centered computing, Visual analytics, Knowledge representation and reasoning, Generative AI.}

\graphicspath{{assets/figures/}{figures/}{pictures/}{images/}{./}} 

\usepackage{booktabs}                  
\usepackage{comment}                   
\usepackage{amsmath,amssymb}           
\usepackage{mathptmx}                  

\usepackage{cuted}
\usepackage{caption}
\AtBeginDocument{\hypersetup{pdfsubject={Preprint}}}

\newcommand{\sparagraph}[1]{\refstepcounter{paragraph}\noindent\textbf{#1\ ---}\label{par:\theparagraph}}

\newcommand{\metric}[2]{%
  \texttt{#1}\,
  \raisebox{-0.15\height}{\includegraphics[height=0.95em]{assets/icons/#2.pdf}}%
}

\begin{document}


\firstsection{Motivation}
\label{sec:introduction}

\maketitle


Data visualization supports understanding and communication across science, business, and public discourse~\cite{schuster_being_2024}. Creating effective visualizations, however, requires nuanced knowledge of perceptual principles, design conventions, and the situated contexts in which audiences interpret data~\cite{kim_how_2025,knoll_gulf_2025,koesten_encountering_2025}. Practitioners who lack formal training in these areas may produce charts that confuse or mislead viewers~\cite{lo_why_2024}. A persistent gap exists between visualization design guidelines and their application in practice~\cite{moritz_formalizing_2019,kim_bridging_2024}.

To close this gap, researchers have developed two types of automated support: \emph{recommendation systems} that generate chart specifications~\cite{mackinlay_automating_1986,wongsuphasawat_voyager_2016,zeng_systematic_2024,dibia_lida_2023,tian_chartgpt_2025} and \emph{feedback systems} that critique existing designs~\cite{mcnutt_linting_2018,chen_vizlinter_2022,hopkins_visualint_2020,shin_visualizationary_2025}. Both depend on codified design knowledge, yet current approaches represent only a narrow slice of what practitioners know. Rule-based systems encode perceptual findings from controlled experiments, but accessibility principles, rhetorical considerations, and editorial heuristics rarely appear in these formalizations~\cite{mcnutt_mixing_2025}. Large Language Models~(LLMs) can reason about such qualities, but without explicit grounding, their recommendations diverge from empirically validated best practices~\cite{wang_dracogpt_2025}. We propose a cataloging scheme that structures visualization design knowledge as natural-language guidelines with typed metadata, capturing the perceptual, contextual, and rhetorical dimensions of reasoning that formal systems cannot express while providing the grounding that generative models lack.

Rule-based systems encode design knowledge as logical constraints~\cite{moritz_formalizing_2019,chen_vizlinter_2022,mcnutt_linting_2018}. These formalisms can weigh competing objectives and guarantee consistent recommendations, but they struggle with design qualities that resist discrete representation. Accessibility~\cite{elavsky_how_2022} or rhetorical tone~\cite{prantl_untangling_2025}, for instance, are not binary properties. Some graphical perception findings, and in certain cases all findings from a study, cannot be expressed within existing recommendation frameworks~\cite{zeng_too_2024}. Rule-based systems couple the definition of good design to the syntax of a particular grammar, such as Vega-Lite~\cite{satyanarayan_vegalite_2017} or Matplotlib~\cite{hunter_matplotlib_2007}. Even for design principles that can be formalized, extending the rule base requires expertise in three areas: the underlying visualization principle, the target grammar in which it must be encoded, and constraint languages like Answer Set Programming~\cite{gebser_multishot_asp_2019}. Few have expertise in all three, restricting who can contribute to the knowledge base.

LLMs offer a flexible alternative. They can produce chart specifications across grammars, decoupling design intent from implementation~\cite{maddigan_chat2vis_2023}. Users might prefer LLM-generated feedback over human-authored advice, citing its breadth, structure, and capacity for rapid iteration~\cite{kim_how_2025,ahn_understanding_2025}. Yet without explicit grounding in design principles, these models reproduce popular but suboptimal patterns~\cite{wang_dracogpt_2025}. Recent work addresses this by incorporating design knowledge into prompts~\cite{shin_visualizationary_2025,choi_bavisitter_2024,gangwar_automated_2025}, improving output quality. However, prompt-embedded knowledge remains opaque, unavailable for inspection, debate, or refinement. The approach also faces a practical constraint. LLM context cannot accommodate the full breadth of visualization research: a chart intended for accessibility review draws on a different set of design guidelines than one optimized for data journalism or scientific communication. Retrieval-Augmented Generation~(RAG) can address this limitation by conditioning model outputs on documents retrieved from an external corpus~\cite{lewis_rag_2020}. For interpretable feedback, however, the underlying knowledge base must be structured for retrieval and be traceable so users can verify the sources of recommendations.

Natural-language sources, such as papers, guidelines, and blog posts, capture situated nuance and contextual exceptions, but they lack the structure necessary for computational use. Formal constraint-based systems support automated reasoning, but their rigid representations cannot express qualities such as rhetorical tone or audience appropriateness, and extending them requires specialized programming skills. Our proposed cataloging scheme combines the strengths of both approaches by structuring visualization design knowledge as natural-language guidelines with typed metadata. The scheme preserves the expressiveness of natural language while introducing the structure that machines require: domain experts author guidelines without programming, systems retrieve them through semantic similarity or categorical filters, and users trace recommendations to their sources. The scheme is grounded in an expert study ($N{=}18$). Using a visualization critique protocol, we observed that experts rarely apply heuristics axiomatically. They invert established rules based on rhetorical goals, such as catchiness or audience literacy, favoring redundancy for novices while penalizing it for experts. To capture this situated reasoning, we model guidelines as structured documents with role-annotated sections. These sections isolate semantically distinct facets of design knowledge: the advice itself, the contexts where it applies, the exceptions that invalidate it, the trade-offs it entails, and the sources from which it derives. Guideline sections embed independently in vector space~\cite{khattab_colbert_2020,vaswani_attention_2017}, supporting deterministic filtering via categorical labels, semantic similarity queries, and structural analysis of the knowledge space itself.

Our cataloging scheme presents a conceptual contribution: a model for representing visualization design knowledge that is machine-queryable, can be authored by experts without programming, is grammar-agnostic, context-sensitive, and traceable to sources. We evaluate this model by instantiating it and demonstrating three analyses that test its core properties. To validate its expressiveness, we encode knowledge from sources that differ in methodology and evidence type. These sources include: quantitative experimental findings from cognitive science~\cite{franconeri_science_2021}, systematic reviews of graphical perception studies~\cite{zeng_review_2023}, normative accessibility criteria~\cite{elavsky_how_2022}, practitioner heuristics from data journalism~\cite{datawrapper_dos_and_donts}, and research on rhetorical aspects of visual communication~\cite{talking_charts_guidelines}. This yields a catalog of 781 guidelines. We then demonstrate how embedding-based analysis of this catalog reveals latent structure in the knowledge space, including conceptual parallels across distant domains, conflicting advice that indicates genuine trade-offs, and boundary conditions where one guideline yields to another. Finally, we demonstrate how the catalog supports grounded feedback from generative models, complementing constraint-based tools like Draco~\cite{moritz_formalizing_2019,yang_draco_2023} and VizLinter~\cite{chen_vizlinter_2022} with situated guidance that users can trace to cited sources.


\vspace{-0.5em}
\section{Related Work}
\label{sec:related-work}

We situate our cataloging scheme among existing approaches to automated visualization design. We first outline requirements for visualization design knowledge representation, drawing on Knowledge-Assisted Visualization~(KAV) and recent findings on situated interpretation. We then evaluate current approaches against these requirements to identify limitations that motivate our work.

\vspace{-0.5em}
\subsection{Requirements for Design Knowledge Representation}
\label{sec:requirements}

What would a representation need to support both human experts and automated systems? We derive five requirements from foundational work on Knowledge-Assisted Visualization~(KAV) and recent empirical studies on how audiences interpret charts.

\textbf{Machine-Readable Encoding.} Chen et al.~\cite{chen_data_2009} define KAV as relying on ``explicit machine-readable knowledge'' to automate processes or guide users. Descriptive text alone is insufficient; to support retrieval and reasoning, design knowledge must be encoded in a form that programs can parse and query~\cite{miksch_knowledgeassisted_2020}.

\textbf{Authorability.} Visualization practice evolves, requiring a representation that supports continuously capturing new expertise. Miksch et al.~\cite{miksch_knowledgeassisted_2020} identify ``externalization'' as a critical process in which practitioners transform tacit knowledge into explicit rules. A useful representation must therefore lower the barrier to contribution, allowing domain experts without extensive technical skills to add and refine guidelines as conventions shift.

\textbf{Separation of Semantics and Syntax.} To ensure portability, the representation must decouple what constitutes good design from a specific toolkit. Miksch et al.~\cite{miksch_knowledgeassisted_2020} argue for ontological approaches that map domain concepts to visual representations while isolating implementation details. This separation ensures that design knowledge remains stable even as rendering engines or grammars change.

\textbf{Situatedness.} Universal design rules often fail in specific contexts. Miksch et al.~\cite{miksch_knowledgeassisted_2020} argue that effective guidance depends on factors beyond the dataset itself, including interaction history and domain conventions. Recent empirical work reinforces this. Knoll et al.~\cite{knoll_gulf_2025} document a ``Gulf of Interpretation'' between chart producers' intent and consumers' understanding, observing that professional designers often rely on conventions that fail to reach specific audience segments. Koesten et al.~\cite{koesten_encountering_2025} highlight how trust and emotional response shape engagement with visualizations. Consequently, design quality cannot be defined universally; it must be specified in relation to specific tasks, audiences, and communicative goals.

\textbf{Traceability.} Finally, recommendations must not operate as a black box. Provenance research in visualization argues that systems should capture not only interactions and insights but also the rationale behind decisions, supporting recall, replication, and validation of prior analyses~\cite{raganCharacterizingProvenanceVisualization2016}. Complementarily, recent work on knowledge-assisted visualization advocates for ``guideline-based'' approaches in which automated encoding decisions are linked to verifiable evidence~\cite{zeng_review_2023,miksch_knowledgeassisted_2020}.

\vspace{-0.5em}
\subsection{Current Approaches to Encoding Design Knowledge}
\label{sec:current-approaches}

We evaluate existing approaches---spanning unstructured natural language, theoretical frameworks, symbolic systems, supervised learning, and generative approaches---against these requirements.

\vspace{.25em}
\sparagraph{Unstructured Natural Language}
Most visualization design knowledge exists as prose scattered across hundreds of sources~\cite{kim_bridging_2024}. Scientific papers report findings on perception and cognition, while practitioner resources include editorial checklists~\cite{datawrapper_dos_and_donts, datavizstyleguide} and discussion forums like VisGuides~\cite{diehl_visguides_2018}, where designers critique real-world charts. Writing in natural language requires no technical expertise, lowering barriers for domain experts to contribute. These resources often capture situated nuance: describing when advice applies, when it doesn't, and why. Natural language also accommodates qualities that resist binary encoding; accessibility guidelines, such as Chartability~\cite{elavsky_how_2022}, pose open-ended questions that guide judgment rather than enforce compliance.
However, scattered knowledge in inconsistent formats remains difficult to use reliably. Relevant guidance may appear in a journal article, a blog post, a paper appendix, or a forum reply, with no shared vocabulary linking related advice across sources. Without structure, there is no systematic way to retrieve guidelines matching a user's specific context, detect when sources offer conflicting recommendations, or trace a suggestion to its original evidence.

\vspace{.25em}
\sparagraph{Theoretical and Domain-Specific Frameworks}
Researchers have also sought to structure design knowledge through theoretical abstractions. Kindlmann and Scheidegger~\cite{kindlmann_algebraic_2014} proposed an algebraic model that defines design validity in terms of mathematical invariants. While rigorous, this approach abstracts away the situated qualities of design, such as rhetorical tone or audience literacy, that practitioners routinely navigate. Choi et al.~\cite{choi_toward_2021} and Kim et al.~\cite{kim_bridging_2024} proposed structuring guidelines via templates containing attributes such as \emph{Context} and \emph{Problem}. While effective for categorizing advice, this framework remains descriptive and does not provide a machine-actionable representation. Similarly, Sedlmair et al.~\cite{sedlmair_visual_2014} established a conceptual framework to categorize tasks in visual parameter space analysis. While this taxonomy effectively guides human designers, it lacks the computational structure required for automated reasoning. Klaffenboeck et al.~\cite{klaffenboeck_rsvp_2025} later operationalized these rules by encoding them into an authoring system. Yet by embedding knowledge directly into application logic, this approach restricts authoring to those with programming expertise and couples design principles to specific software implementations.
To address the need for decoupled, standardized formats, recent work by Ottley~\cite{ottley2026consensus} systematically analyzed 53 organizational style guides, using a large language model to extract unstructured text into a JSON schema capturing actions, targets, and exceptions. While this approach successfully structures design guidelines without coupling them to specific software, its primary contribution remains a descriptive corpus for analyzing industry consensus. The underlying JSON representation functions as a static snapshot for exploratory data analysis rather than a computational framework. Consequently, it lacks the extensibility and computational structure required to serve as a dynamic, machine-actionable knowledge base for grounding generative reasoning. 

\vspace{.25em}
\sparagraph{Symbolic and Rule-Based Approaches}
To make design rules computable, prior work has encoded them as logical constraints. ``Visualization linting'' validates chart specifications against codified best practices. VizLinter~\cite{chen_vizlinter_2022} and earlier work by McNutt and Kindlmann~\cite{mcnutt_linting_2018} detect structural flaws, while VisuaLint~\cite{hopkins_visualint_2020} provides in-situ annotations during authoring. These methods have been adapted for specific domains, including choropleth validation~\cite{lei_geolinter_2024} and educational grading~\cite{hull_visgrader_2024}.
The most comprehensive formalization known to the authors is Draco~\cite{moritz_formalizing_2019}, which models design knowledge as weighted constraints in Answer Set Programming (ASP)~\cite{gebser_multishot_asp_2019}. The original system was coupled to Vega-Lite~\cite{satyanarayan_vegalite_2017}; Draco~2~\cite{yang_draco_2023} introduced a renderer-agnostic schema, decoupling design principles from the specific grammar used to implement charts. Kim and Heer~\cite{kim_data_2026} extended this work by developing data augmentation techniques to improve knowledge base coverage, generating novel chart pairs through design permutations, and scaling labeling efforts to learn updated feature weights. KG4Vis~\cite{li_kg4vis_2022} encodes knowledge as a graph that links data features to visualization design choices; embeddings learned from this graph enable recommendation and rule extraction.
These symbolic approaches satisfy machine-readability and, in the case of Draco~2, separation of semantics from syntax. However, they struggle with situatedness. A constraint can verify that a chart has a title, but it cannot assess whether that title's tone matches the creator's rhetorical intent or suits the target audience. Such qualities are subjective and cannot be reduced to binary predicates. Maintaining these formalisms also requires technical expertise, such as ASP, which limits who can contribute.

\vspace{.25em}
\sparagraph{Supervised Learning}
VizML~\cite{hu_vizml_2019} and Data2Vis~\cite{dibia_data2vis_2019} learn mappings from dataset characteristics to visual specifications, training on publicly available chart corpora. These systems demonstrate that design patterns can be learned from examples, but they face two limitations. First, they optimize for frequency over quality, thereby risking the perpetuation of common pitfalls, such as ineffective 3D charts. Second, their reasoning is opaque; they cannot explain why a particular design was chosen, failing the traceability requirement.

\vspace{.25em}
\sparagraph{Generative Approaches}
LLMs enabled new approaches to visualization authoring. LIDA~\cite{dibia_lida_2023} and Chat2VIS~\cite{maddigan_chat2vis_2023} translate natural language queries to visualization code, while ChartGPT~\cite{tian_chartgpt_2025} uses chain-of-thought reasoning to decompose user goals into step-by-step specifications. These systems can separate the abstract concept of a desired chart from the concrete implementation details of rendering, but the design knowledge guiding their outputs, such as why a bar chart suits a particular data distribution, remains implicit in the model weights.
Recent empirical work has investigated LLMs for visualization design tasks. Kim et al.~\cite{kim_how_2025} compared ChatGPT-generated responses with human replies to design questions on the VisGuides forum and conducted a user study where practitioners received feedback from both sources. Participants valued ChatGPT's ability to generate diverse design options rapidly and appreciated the structured, comprehensive nature of its responses. A follow-up analysis by Ahn and Kim~\cite{ahn_understanding_2025} examined why users preferred LLM-generated advice, finding that ChatGPT outputs exhibited broader coverage, more consistent rhetorical structure, and stronger emphasis on task-oriented feedback than human responses. These findings suggest that LLMs possess properties (fluency, breadth, and responsiveness) that practitioners find valuable in design support tools.
However, the studies identified limitations. LLMs struggled with nuanced contextual understanding, sometimes offering generic advice that failed to account for specific audience or domain constraints~\cite{kim_how_2025}. DracoGPT~\cite{wang_dracogpt_2025} probed the design preferences encoded in several LLMs and found they substantially diverge from guidelines derived from human subjects experiments, suggesting these models favor popular conventions over empirically effective designs. Without explicit grounding, these systems cannot explain their choices with verifiable principles, nor can their preferences be audited or updated without retraining or finetuning.
Recent work attempts to ground generative models through in-context learning, where examples or instructions provided in the prompt steer the model's output without updating its weights. Bavisitter~\cite{choi_bavisitter_2024} uses rule-based heuristics to detect violations in generated specifications, then prompts the model to resolve them. Visualizationary~\cite{shin_visualizationary_2025} extracts perceptual features (saliency, color contrast) from rendered images and injects them into prompts alongside design best practices, enabling critiques tailored to the specific chart instead of relying on generic patterns. These methods produce feedback that responds to specific design contexts, but they tend to fragment the knowledge landscape. The heuristics in Bavisitter are embedded in its detection module, while the critique logic in Visualizationary is within the prompt templates. While effective for their intended applications, this knowledge is difficult to inspect, reuse across systems, or refine through community debate.

\vspace{-0.5em}
\section{Formative Study on Situated Reasoning}
\label{sec:formative-study}

To inform the design of our cataloging scheme, we conducted an observational study with visualization experts ($N=18$). Prior taxonomies have classified guidelines by chart type or task~\cite{choi_toward_2021}; our objective was instead to characterize how experts reason about design in context: how they adapt heuristics to specific situations, how they prioritize when guidelines conflict, and when they reject established advice entirely.

\vspace{-0.5em}
\subsection{Methodology}
\label{sec:study-methodology}

We recruited 18 participants at the IEEE VIS 2025 conference in Vienna. The cohort spanned academia (12), industry (3), and non-profit or government organizations (3), with experience levels ranging from early-career researchers to senior experts with over 20 years of practice. Participants are identified herein as \texttt{P01}--\texttt{P18}.

\textbf{Stimuli and Protocol.} We employed a critique-and-meta-critique protocol centered on three real-world design scenarios adapted from the \emph{Datawrapper} ``Fix My Chart'' series~\cite{datawrapper_dos_and_donts}. The scenarios covered common chart types while introducing distinct encoding challenges: (1)~small multiple donut charts showing bridge conditions across U.S. states, where experts often recommend bar charts or tables for improved precision~\cite{datawrapper_fmc_bar}; (2)~an OECD employment line chart with challenges around indexing and baseline manipulation~\cite{datawrapper_fmc_line}; and (3)~a choropleth map requiring participants to navigate trade-offs between geographical fidelity, legend complexity, and color perception~\cite{datawrapper_fmc_choropleth}.
The protocol proceeded in three stages. First, participants reviewed the original chart and the creator's stated intent, providing unprompted verbal feedback. Second, in a \emph{meta-critique} phase, they evaluated discrete improvement suggestions drawn from the original \emph{Datawrapper} expert analysis. We asked participants not only whether they agreed, but why a technically defensible suggestion might be inappropriate for the specific case. Third, we probed situatedness by asking how their advice would differ for three distinct audiences: an expert who uses dashboards daily, a casual consumer who encounters charts occasionally, and a novice with limited experience reading visualizations.

\textbf{Data Collection.} Sessions lasted approximately 15 minutes and were audio-recorded with written informed consent. Participants received a confectionery item as compensation. Audio was transcribed using a locally hosted speech-to-text model; recordings were deleted after transcript verification.

\vspace{-0.5em}
\subsection{Findings}
\label{sec:study-findings}

We organize our findings into five themes, each of which motivates a design decision in the cataloging scheme.

\textbf{Audience-Dependent Advice.} Experts routinely adapted recommendations based on audience and medium. \texttt{P13} recommended maps for casual readers but tables for policymakers; \texttt{P16} noted that chart type ``really doesn't matter'' for experts who ``will navigate through it'', whereas novices benefit from familiar conventions. \texttt{P14} distinguished interactive from static contexts. Guidelines cannot simply prescribe a chart type; they must specify the conditions under which the advice applies, motivating the inclusion of context and exception fields.

\textbf{Rationale and Trade-offs.} Participants articulated mechanisms instead of binary judgments. \texttt{P11} explained that novices ``reduce it to what you understand, which is colors'', justifying why palette choice matters. \texttt{P14} described a title as ``more catchy'' and ``timely''; \texttt{P08} argued that simplification ``hides it a lot more''. These rhetorical and perceptual qualities resist Boolean encoding. The scheme must preserve rationale as natural language, separate from the instruction itself.

\textbf{Disagreement and Provenance.} Participants frequently rejected suggestions that Datawrapper's experts had offered as improvements. \texttt{P11} dismissed a narrative annotation as ``very confusing''; \texttt{P12} questioned y-axis indexing; \texttt{P15} criticized a star rating for lacking transparency. These disagreements reflect different values, not errors. The scheme must track provenance so users can assess credibility and recognize when guidelines represent one perspective among several.

\textbf{Conditional Inversion.} Experts sometimes inverted standard heuristics. \texttt{P11} noted that redundancy (typically discouraged) helps low-literacy audiences. \texttt{P13} observed that searchable interfaces aid through exploration but hinder overview-first reading for casual consumers. \texttt{P14} suggested that annotations intended to help casual consumers might confuse experts who expect interactivity. Advice and its negation can both be valid; the scheme must support explicit exception conditions.

\textbf{Domain and Task Specificity.} Feedback depended on analytical purpose, not just chart type. \texttt{P11} noted that labeling places of interest is ``very domain-specific, very contextual''. \texttt{P10} emphasized understanding ``which kind of data they would like to showcase''. Existing systems retrieve visual forms by data type; the scheme must express factors like domain and task as metadata, enabling retrieval by situated analytical purpose.


\vspace{-0.5em}
\section{The Cataloging Scheme}
\label{sec:representation}

The requirements outlined in \cref{sec:requirements} and the findings presented in \cref{sec:study-findings} jointly inform the scheme's structure. Experts often express guidance conditionally, but the qualities they evaluate are not always reducible to Boolean checks. A system can encode when advice should apply, yet it is harder to formalize open-ended judgments such as whether a design is easy to understand for a particular audience, rhetorically appropriate, or engaging. Experts also adapt advice to audience and medium and invert heuristics under specific conditions. To capture these properties while ensuring machine-readability, low-barrier authoring, and traceability, we model guidelines as documents with role-annotated sections and categorical metadata, allowing each section to be analyzed independently.

\vspace{-0.5em}
\subsection{The Guideline Model}

\begin{figure}[!b]
\centering
\vspace{-1.5em}
\caption{Structure of a guideline in the cataloging scheme. The tuple $g = (\texttt{id}, \texttt{title}, \texttt{description}, \mathcal{L}, \mathcal{S}, \mathcal{R})$ organizes design knowledge into metadata for identification, categorical labels for deterministic filtering, role-annotated sections that embed independently in vector space, and bibliographic references for provenance.}
\label{fig:guideline-model}
\includegraphics[width=\columnwidth]{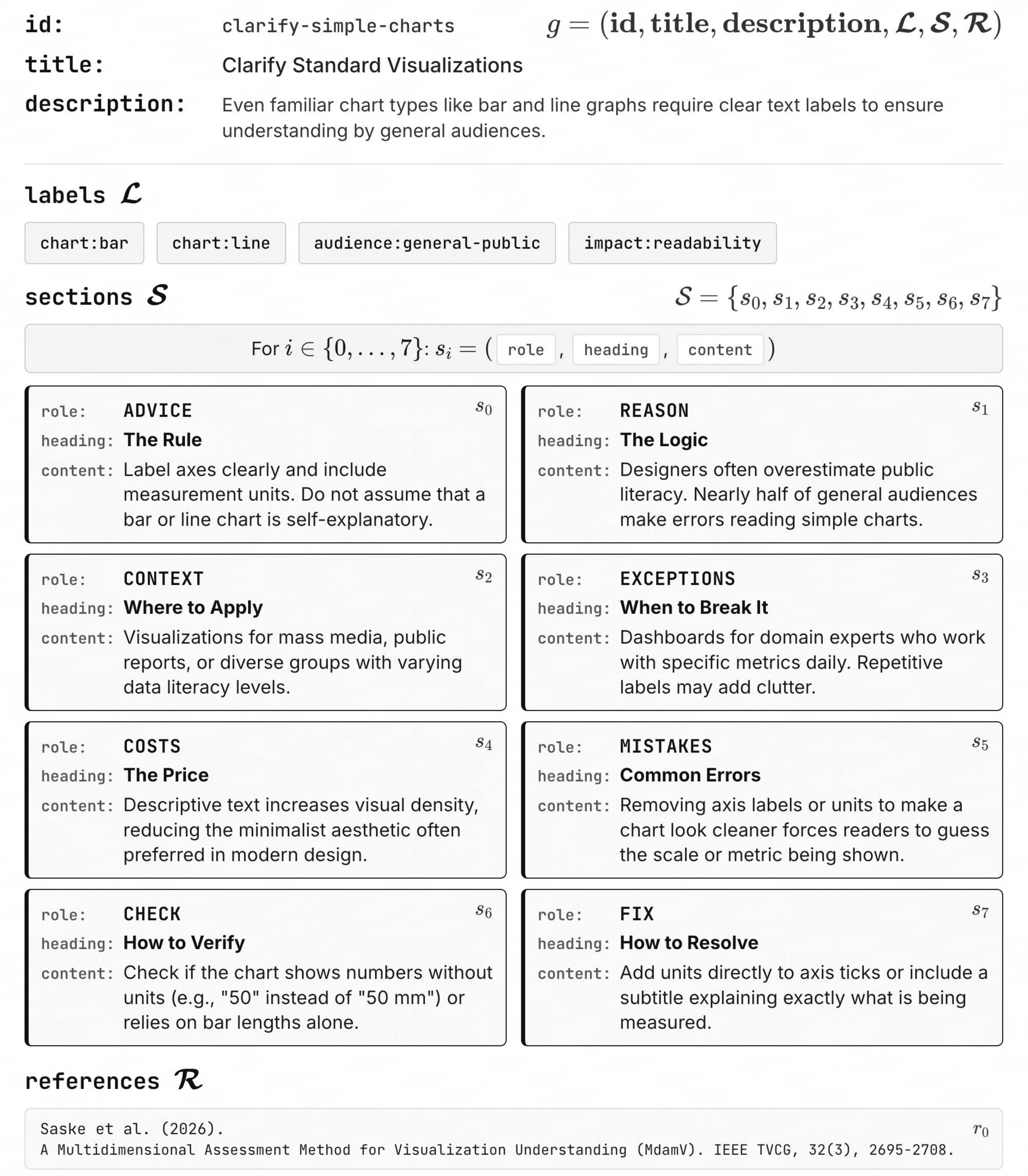}
\vspace{-1.5em}
\end{figure}

We model a guideline $g$ as the tuple:
\begin{equation}
\label{eq:guideline}
g = (\texttt{id}, \texttt{title}, \texttt{description}, \mathcal{L}, \mathcal{S}, \mathcal{R})
\end{equation}
where \texttt{id} is a unique identifier; \texttt{title} is an imperative statement of the advice; \texttt{description} is a brief summary; $\mathcal{L}$ is a set of categorical labels; $\mathcal{S}$ is an ordered sequence of role-annotated sections; and $\mathcal{R}$ is a set of bibliographic references. \cref{fig:guideline-model} illustrates this structure with a concrete example.

\vspace{.25em}
\sparagraph{Taxonomic Labels}
The label set $\mathcal{L}$ provides a vocabulary for categorical filtering. Each label follows a \texttt{category:value} structure. The default categories were informed by the formative study: participants reasoned along dimensions of chart type, analytical task, audience characteristics, and design goals, yielding categories such as \texttt{chart:bar}, \texttt{task:comparison}, and \texttt{audience:novice}.

The scheme imposes no fixed vocabulary. Authors may introduce arbitrary categories (e.g., \texttt{domain:clinical}, \texttt{tool:tableau}, \texttt{source:peer-reviewed}) without modifying the underlying model. This extensibility enables communities to develop specialized vocabularies while maintaining interoperability with shared tooling. The same explicit label structure also makes conceptually overlapping labels visible, which creates an entry point for later consolidation through human review or concept-induction methods~\cite{lamConceptInductionAnalyzing2024}.

\vspace{.25em}
\sparagraph{Section Roles}
Each section $s \in \mathcal{S}$ is a triple:
\begin{equation}
\label{eq:section}
s = (\texttt{role}, \texttt{heading}, \texttt{content})
\end{equation}
where \texttt{role} declares the section's function, \texttt{heading} provides a human-readable title, and \texttt{content} contains the prose. The role isolates a semantically distinct aspect of design reasoning; this isolation has computational consequences that we formalize in \cref{sec:analysis}.

\noindent The default role taxonomy reflects patterns observed in our study:

\begin{description}[leftmargin=2.8cm, style=sameline, noitemsep, topsep=2pt]
  \item[\texttt{advice}] The instruction itself.
  \item[\texttt{reason}] The mechanism explaining why the advice holds.
  \item[\texttt{context}] Preconditions under which the advice applies.
  \item[\texttt{exceptions}] Situations that invalidate the advice.
  \item[\texttt{costs}] Trade-offs incurred by following the advice.
  \item[\texttt{mistakes}] Common anti-patterns.
  \item[\texttt{check}] Diagnostic criteria for detecting violations.
  \item[\texttt{fix}] Remediation strategies.
\end{description}

This taxonomy is extensible. A community focused on accessibility might add roles for \texttt{assistive-technology} or \texttt{cognitive-load}; a team developing educational materials might introduce \texttt{learning-objective} or \texttt{prerequisite}. The formal model accommodates such extensions: any string may serve as a role identifier, and the analytical operators defined below apply uniformly regardless of which roles a catalog employs.

\vspace{.25em}
\sparagraph{Provenance}
The reference set $\mathcal{R}$ links each guideline to its sources. By maintaining explicit provenance, recommendations can be attributed to specific studies or authors. Users can then assess credibility based on origin, a property the formative study identified as important when experts disagreed on best practices (\cref{sec:study-findings}).

\vspace{-0.5em}
\subsection{The Catalog}
\label{sec:catalog}

A \emph{catalog} $\mathcal{C}$ is a finite set of guidelines \( \mathcal{C} = \{g_1, g_2, \ldots, g_N\} \) where each $g_i$ conforms to the structure in \cref{eq:guideline}. The operators defined in \cref{sec:analysis} take the catalog as input and produce derived structures (similarity matrices, conflict sets, ranked retrievals) as output.
Because guidelines share a common schema, the catalog admits uniform iteration. For any role $r$, we can extract the set of all sections with that role: \( \mathcal{S}_r = \{s \mid g_i \in \mathcal{C},\; s \in \mathcal{S}_i,\; s.\texttt{role} = r\} \). This projection enables operations that compare guidelines along a single dimension. For example, extracting all \texttt{context} sections enables the identification of guidelines addressing similar situations, regardless of whether their advice or rationale differs.

\vspace{-0.5em}
\subsection{Embedding Sections for Retrieval and Analysis}
\label{sec:embedding}

Natural language preserves expressiveness but does not support similarity computation or structured retrieval. To enable these operations, we embed the section text into a continuous vector space. This approach builds on representation learning, which has established that embedding functions can map semantically related texts to nearby points in high-dimensional space~\cite{bengio_representation_2013}, and on sentence embedding methods that produce vectors where cosine similarity correlates with semantic similarity~\cite{reimers_sentencebert_2019}.
Let $\mathcal{T}$ denote the space of natural-language text strings. Let $\Phi: \mathcal{T} \rightarrow \mathbb{R}^d$ denote an embedding function mapping text to a $d$-dimensional vector. For a guideline $g_i$ with sections $\mathcal{S}_i$, we compute \( \vec{v}_{i,r} = \Phi(s_{i,r}.\texttt{content}) \) where $s_{i,r}$ is the section of $g_i$ with role $r$. We embed sections independently rather than concatenating all sections into a single guideline vector. This separation allows role-specific queries: a search for similar contexts retrieves guidelines facing comparable situations regardless of whether their advice differs, while a search for similar advice identifies guidelines offering related recommendations regardless of the situations they address.
The choice of $\Phi$ is a parameter of the scheme. Options range from weighted averaging of word vectors~\cite{arora_simple_2017} to transformer-based encoders~\cite{khattab_colbert_2020}. The formal model accommodates any function with the type signature $\mathcal{T} \rightarrow \mathbb{R}^d$; we report a concrete instantiation in \cref{sec:analysis}.
This embedding enables two operations on the catalog. First, cosine similarity between vectors correlates with semantic relatedness~\cite{arora_simple_2017}, enabling nearest-neighbor retrieval: given a user query, we can locate guidelines whose sections are semantically proximate. Second, embedding spaces learned from text exhibit compositional structure, in which vector arithmetic expresses semantic relationships~\cite{mikolov_linguistic_2013,reimers_sentencebert_2019}. Differences and combinations of section vectors can be used to compare guidelines along specific dimensions.
Because sections are embedded by role, the catalog supports comparisons unavailable in representations that treat guidelines as monolithic units. Comparing \texttt{context} vectors across guidelines identifies entries that address similar situations, even when their advice differs. Comparing one guideline's \texttt{advice} vector to another's \texttt{mistakes} vector can reveal cases where one source recommends something, while another explicitly discourages it. Measuring the difference between \texttt{advice} similarity and \texttt{context} similarity across a pair of guidelines can identify transferable principles (similar advice for different situations) or genuine conflicts (divergent advice for similar situations). We demonstrate these operations in \cref{sec:analysis}.

\vspace{-0.5em}
\subsection{Properties of the Scheme}
\label{sec:properties}

We designed the scheme to meet the requirements in \cref{sec:requirements}. We summarize the intended properties below and provide evidence that they hold in practice in \cref{sec:operationalization}.
Section roles and categorical labels support direct parsing, and vector projections enable similarity computation and nearest-neighbor retrieval, satisfying \emph{machine-readable encoding}. Natural language serves as the expressive medium, so authoring a guideline requires structured prose instead of formal logic or programming, satisfying \emph{authorability}. The scheme describes design knowledge independently of visualization grammars. A guideline about color encoding applies whether the target system is Vega-Lite~\cite{satyanarayan_vegalite_2017} or Matplotlib~\cite{hunter_matplotlib_2007}, achieving \emph{separation of semantics and syntax}. Roles such as \texttt{context} and \texttt{exceptions} explicitly model conditions under which advice applies, providing \emph{situatedness}. The reference set $\mathcal{R}$ links each guideline to its sources, ensuring \emph{traceability}.

\vspace{-0.5em}
\section{Implementation of the Scheme}
\label{sec:operationalization}

The previous section formally defined the cataloging scheme. We now demonstrate that it can be operationalized. We first describe a concrete instantiation using Markdown documents, then show that diverse sources can be mapped to the structure, that the operators yield interpretable results, and that this structured knowledge can ground generative reasoning and situated feedback. These demonstrations test the scheme's expressiveness and utility; they are not presented as a production system. Implementation details and interactive notebooks appear in the supplementary materials.

\vspace{-0.5em}
\subsection{Instantiation}
\label{sec:instantiation}

We instantiate the scheme using Markdown documents with YAML frontmatter. Metadata fields (\texttt{id}, \texttt{title}, \texttt{description}, \texttt{labels}, \texttt{bibliography}) appear in the frontmatter. The document body contains sections demarcated by headings with role annotations embedded as HTML comments (e.g., \texttt{<!-- role: advice -->}). This format requires no specialized tooling, integrates with version control systems, and remains human-readable when rendered. A parser extracts the structured representation by matching heading patterns and role annotations.
This instantiation is one of many. The formal model (\cref{eq:guideline}, \cref{eq:section}) could be realized as JSON documents, database records, or knowledge graph entries. Markdown prioritizes authorability and transparency: experts can write and review guidelines using familiar tools, and content remains human-readable without specialized software. Supplementary materials include the full catalog, executable notebooks for each demonstration below, source code, and a narrated video walkthrough of the case studies with timestamps. Notebooks render as static HTML for inspection without running code.

\begin{figure*}[t]
\centering
\includegraphics[width=\textwidth]{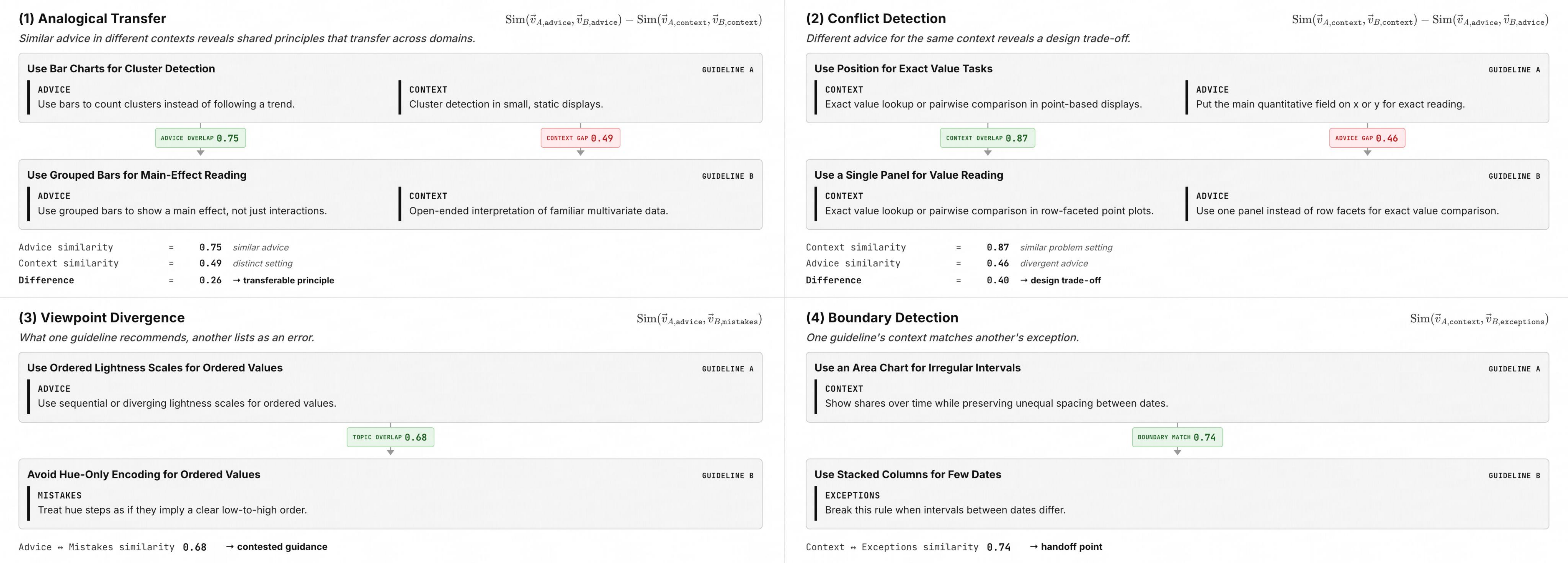}
\vspace{-2em}
\caption{Four operators over section embeddings. Each panel shows two guidelines, the sections compared, similarity scores, and the resulting interpretation. (1) Analogical transfer: high advice similarity with low context similarity indicates a shared principle across domains. (2) Conflict detection: high context similarity with low advice similarity reveals a potential design trade-off. (3) Viewpoint divergence: when one guideline's advice resembles another's listed mistakes, sources might disagree on best practice. (4) Boundary detection: when one guideline's context matches another's exceptions, a handoff condition emerges where one guideline yields to another.}
\vspace{-2em}
\label{fig:operators}
\end{figure*}

\vspace{-0.5em}
\subsection{Cataloging Diverse Knowledge Sources}
\label{sec:cataloging}

A necessary property of the scheme is that it can accommodate knowledge from sources that differ in methodology, evidence type, and rhetorical form. Prior formalizations have focused on perceptual findings from controlled experiments~\cite{moritz_formalizing_2019, zeng_review_2023}, but practitioners also draw on rhetorical considerations, accessibility principles, and editorial heuristics. To test expressiveness, we mapped five source categories to the scheme, selecting them to span the spectrum identified in prior work on the visualization knowledge gap~\cite{kim_bridging_2024, choi_toward_2021}: quantitative experimental findings, systematic reviews, normative accessibility criteria, practitioner heuristics derived from editorial experience, and qualitative research on communicative intent. The source corpus described below instantiates these categories in the catalog.

\textbf{Cognitive Science and Perception.} We processed 106 papers cited in Franconeri et al.'s review ``The Science of Visual Data Communication''~\cite{franconeri_science_2021}, which synthesizes findings on low-level visual processing (saliency, ensemble coding, color discrimination), higher-order cognition (bias, memory, narrative framing), and applied domains (risk communication, uncertainty visualization).

\textbf{Collated Graphical Perception.} Zeng and Battle~\cite{zeng_review_2023} systematically collated findings from 58 empirical studies on graphical perception, extracting performance metrics (accuracy, response time, bias) that rank visual encodings and chart types.

\textbf{Accessibility Standards.} Chartability~\cite{elavsky_how_2022} collates accessibility principles for data visualization through 50 heuristic evaluation criteria derived from inclusive design principles.

\textbf{Practitioner Heuristics.} Datawrapper's ``Dos and Don'ts'' blog series~\cite{datawrapper_dos_and_donts} represents tacit knowledge accumulated through data journalism practice, communicated across 36 posts. These heuristics emphasize clarity, aesthetics, and reader engagement---knowledge that rarely appears in formal systems yet reflects lessons from producing thousands of charts for general audiences.

\textbf{Rhetoric Guidelines.} The Talking Charts project~\cite{talking_charts_guidelines} produced 32 findings on how producers and consumers interpret visualizations, addressing rhetoric, resonance, and audience connection---aspects that perception-focused formalizations typically omit.

\cref{tab:source-stats} summarizes the resulting catalog. By incorporating rhetorical guidelines and practitioner heuristics alongside perception research, the catalog captures knowledge that existing formal systems struggle to represent.

\begin{table}[t]
\centering
\caption{Guidelines cataloged by source.}
\label{tab:source-stats}
\begin{tabular}{@{}lr@{}}
\toprule
\textbf{Source Category} & \textbf{Guidelines} \\
\midrule
Cognitive Science and Perception & 366 \\
Practitioner Heuristics & 206 \\
Collated Graphical Perception & 127 \\
Accessibility Standards & 50 \\
Rhetoric Guidelines & 32 \\
\midrule
\textbf{Total} & 781 \\
\bottomrule
\end{tabular}
\end{table}

We used a generative model~(GPT 5.4) to extract guidelines from source texts and map them to the schema's fields, preserving BibTeX keys for provenance. This treats the model as a restructuring tool rather than a knowledge generator~\cite{ke_large_2026}: the model reformulates existing prose into the scheme's structure while leaving the original sources inspectable. Provenance is maintained through the reference set $\mathcal{R}$, ensuring that each guideline links to its source. The quality of this particular catalog is not our claim; the point is that the scheme supports such extraction workflows whose outputs remain inspectable, source-linked, and reusable in downstream retrieval and evaluation.

\vspace{-0.5em}
\subsection{Structural Analysis of the Knowledge Space}
\label{sec:analysis}

We embedded each section using a transformer-based encoder (OpenAI \texttt{text-embedding-3-large}, $d=3072$) and indexed the resulting vectors with a Hierarchical Navigable Small World~(HNSW) index~\cite{malkov_efficient_2020} to support approximate nearest-neighbor search. For analysis, each of the $781$ guidelines contributes $10$ embeddings: one for the full guideline document, one overview embedding constructed from the title and description, and one embedding for each of the eight role-specific sections. \cref{fig:guideline-embedding-atlas} shows the resulting embedding space over all $7{,}810$ vectors using Embedding Atlas~\cite{ren_embedding_2025}. The analyses below execute as SQL queries over this structure; implementation details appear in the supplementary materials.
We report four operators that exploit the geometric properties of section embeddings to reveal structure in the knowledge space (\cref{fig:operators}). Each compares guideline pairs along specific dimensions using cosine similarity, denoted $\text{Sim}(\cdot, \cdot)$.
These operators expose structure in the catalog, but they do not equate semantic proximity with direct applicability. A guideline can be close to a query because it discusses the same chart type while arguing against its use. We therefore treat embedding-based operators as tools for analysis and candidate retrieval, not as self-sufficient evidence that the retrieved guidance is correct for a given design problem.

\textbf{Analogical Transfer.} Guidelines may share underlying principles despite addressing unrelated domains. To surface such cases, we search for pairs where \texttt{advice} vectors are similar but \texttt{context} vectors are distant: \( \text{Sim}(\vec{v}_{A,\texttt{advice}}, \vec{v}_{B,\texttt{advice}}) - \text{Sim}(\vec{v}_{A,\texttt{context}}, \vec{v}_{B,\texttt{context}}) \).

Panel~(1) of \cref{fig:operators} pairs a guideline that recommends bars for cluster detection with one that recommends grouped bars for reading main effects. The settings differ: one concerns small, static displays, whereas the other concerns open-ended interpretation of familiar multivariate data. However, both guidelines use bar-based encodings to support grouped comparison over trend reading. The operator therefore surfaces a principle that transfers across settings.

\textbf{Conflict Detection.} Guidelines can also address the same problem setting through different interventions. We search for pairs with similar contexts but less similar advice: \( \text{Sim}(\vec{v}_{A,\texttt{context}}, \vec{v}_{B,\texttt{context}}) - \text{Sim}(\vec{v}_{A,\texttt{advice}}, \vec{v}_{B,\texttt{advice}}) \).

Panel~(2) of \cref{fig:operators} pairs two guidelines aimed at exact value lookup and pairwise comparison in point-based displays. One emphasizes encoding: place the primary quantitative field on a positional axis. The other emphasizes layout: avoid row facets when exact comparison is required. The operator therefore surfaces alternative interventions for the same reading task.

\textbf{Viewpoint Divergence.} Some recommendations are contested across the field. To identify such cases, we measure similarity between one guideline's \texttt{advice} and another's \texttt{mistakes}: \( \text{Sim}(\vec{v}_{A,\texttt{advice}}, \vec{v}_{B,\texttt{mistakes}}) \).

Panel~(3) of \cref{fig:operators} pairs a guideline that recommends sequential or diverging lightness scales for ordered values with one that treats hue steps as if they implied a clear low-to-high order as a mistake. The operator therefore surfaces viewpoint divergence because the same underlying design principle is encoded in opposite rhetorical forms: as recommended practice in one guideline and as an error condition in another.

\textbf{Boundary Detection.} Guidelines often have exceptions that indicate when alternative advice should take precedence. To propose such handoffs, we measure similarity between one guideline's \texttt{context} and another's \texttt{exceptions}: \( \text{Sim}(\vec{v}_{A,\texttt{context}}, \vec{v}_{B,\texttt{exceptions}}) \).

Panel~(4) of \cref{fig:operators} pairs a guideline that recommends area charts when shares over time must preserve unequal spacing between dates with one that recommends stacked columns when the number of dates is small. The second guideline explicitly lists irregular temporal intervals as an exception. The operator therefore surfaces a handoff condition: stacked columns suit a small number of evenly spaced dates, whereas area charts become preferable when unequal spacing carries meaning.

\vspace{-0.5em}
\subsection{Grounding Generative Feedback}
\label{sec:application}

Finally, we demonstrate that our cataloging scheme can ground feedback from generative models and complement symbolic validation with situated guidance in a concrete design scenario.

\vspace{.25em}
\sparagraph{Scenario}
A communicator is adapting a bar chart from a city drought report for a grocery flyer aimed at general-public shoppers. Readers will likely see the chart briefly while deciding what protein to buy for dinner. The goal is to help someone considering beef quickly compare it with lower-water alternatives such as chicken, eggs, or pulses. The chart retains a horizontal bar layout sorted from highest to lowest water use so beef stands out, but the designer is unsure whether rank order best supports this substitute-comparison task.
This scenario tests whether automated feedback can address outlet-specific and task-specific concerns that symbolic linters cannot assess. The chart is structurally sound: position encodes quantity, axes are labeled, and no expressiveness violations occur. The question is whether a report-style ranking also supports rapid substitute comparison in a brief flyer-reading context.

\vspace{.25em}
\sparagraph{Feedback Pipeline}
The feedback pipeline combines pixel-level extraction, symbolic validation, and catalog-grounded retrieval. DePlot~\cite{liu_deplot_2023} recovers the underlying data from the rasterized image. VizLinter~\cite{chen_vizlinter_2022} and Draco~\cite{yang_draco_2023} then check the inferred specification against structural rules and perceptual constraints. For our scenario, both tools report no violations.
Symbolic tools confirm structural validity but cannot assess whether the report-style ordering supports quick substitute comparison in a flyer. The key enabler here is the catalog's structure: because guidelines are stored as labeled, role-annotated records, a Reasoning and Acting (ReAct) agent~\cite{yao_react_2023} can query them directly. The agent first inspects the schema and vocabulary via SQL, maps the situation to a small set of labels such as \texttt{audience:general-public}, \texttt{task:compare}, and \texttt{lever:layout-structure}, then shortlists candidates and reads the relevant sections for producing grounded feedback.

\vspace{.25em}
\sparagraph{Effect of Grounding}
We compare three configurations. An off-the-shelf LLM without catalog access produces fluent but generic advice without source citations. Adding linting results provides technical context but does not resolve the grounding problem: feedback combines valid structural observations with unverifiable claims about audience needs.
The catalog-grounded pipeline retrieves guidelines for spatial grouping in comparison tasks, minimizing visual distance for older viewers, and adapting complexity to target audiences. Feedback notes that strict descending sort pushes substitute proteins far apart and suggests category-based grouping to reduce visual search load. Each suggestion cites a guideline so users can inspect the rationale, exceptions, and primary sources.
This demonstration illustrates three properties enabled by the cataloging scheme: \emph{situatedness} (the agent adapts feedback to the stated audience, outlet, and task), \emph{traceability} (suggestions cite verifiable guidelines), and \emph{extensibility} (guidance comes from a catalog that experts can update as knowledge evolves).

\vspace{-0.5em}
\section{Empirical Evaluation of the Scheme}
\label{sec:evaluation}

We evaluate how the cataloging scheme supports generative models in visualization design reasoning. We use VisEval~\cite{chenVisEvalBenchmarkData2025} as the source of realistic natural-language visualization queries and VisJudge-7B from VisJudgeBench~\cite{xieVisJudgeBenchAestheticsQuality2026} for automatic multi-dimensional judgments that approximate expert assessment of visualization quality. Our focus is on whether retrieved design knowledge shapes the designs that models produce and whether those designs remain stable across different models and chart grammars.

\vspace{.25em}
\sparagraph{Experimental Design}
We run the experiment with three models of the GPT-family chosen to span distinct capability profiles while controlling for the model family: GPT-5.4, a frontier general-purpose foundation model; GPT-5.3 Codex, a frontier coding-specialized model; and GPT-5.4-mini, a smaller efficient model. This choice allows us to test whether the effects of catalog grounding persist across differences in model purpose and reasoning profile rather than reflecting vendor-specific differences. The chart grammars are Matplotlib~\cite{hunter_matplotlib_2007}, an imperative API; Altair~\cite{VanderPlas2018}, a declarative interface to Vega-Lite~\cite{satyanarayan_vegalite_2017}; and Plotly~\cite{plotly}, a figure-construction API for interactive layered charts. Together, they span distinct authoring styles, and let us examine how the same analytical query is realized across grammars.
We study two settings. In \texttt{refine}, the natural-language request already prescribes a high-level design. A query such as ``Show the number of films for each type in a pie chart.'' fixes the chart family in advance, so the model's responsibility is to implement that design in the chosen grammar and reason about refinement choices such as labeling, ordering, annotation, and other grammar-specific details. In \texttt{select}, the request specifies only analytical intent. A query such as ``Show the number of films by type.'' leaves the chart family open, so the model is responsible for both selecting an appropriate design and implementing it in the chosen grammar. For each query, we compare an ungrounded condition, in which the model relies on its embedded visualization knowledge, with a grounded condition, in which guidance is retrieved through our cataloging scheme. This design allows us to assess how structured guidance affects design quality and chart-choice consistency across models and chart grammars.

\vspace{.25em}
\sparagraph{Adapting VisEval for the Experiment}
We build the controlled experiment on VisEval~\cite{chenVisEvalBenchmarkData2025}, a benchmark for evaluating natural-language-to-visualization (NL2VIS) generation methods. VisEval contains $2,524$ natural-language query and visualization pairs over $1,150$ unique visualizations, spanning $146$ databases and $7$ chart families. This breadth makes it a strong basis for studying how generative models respond to realistic visualization queries. In its original form, each query already prescribes a chart family, so the benchmark aligns naturally with the \texttt{refine} setting introduced above. To support \texttt{select} as well, we extend VisEval with chart-agnostic queries that preserve analytical intent while leaving chart choice open. The same dataset can then support both settings in our experiment.
We adapt the benchmark in two ways. First, we improve the dataset's executability to support our generation study.\footnote{In working with the released benchmark, we found issues such as missing table files and syntactic and semantic SQL errors. Our supplementary material details these issues and our repair procedure.} Second, we canonicalize the original chart-prescriptive queries into chart-type-agnostic requests. This step applies to $2,518$ unique raw query strings associated with $1,150$ target visualizations. It yields $1,125$ canonical intents because some distinct target visualizations differ only in chart family while expressing the same underlying analytical goal. For example, pie-chart and bar-chart requests about the number of students advised by each faculty rank collapse to the same canonical intent. We generate these rewrites with GPT-5.4, check them with a second LLM judge for consistency, and manually spot-check the outputs to confirm that chart-type references are removed while analytical intent is preserved.
We further enrich each canonical query with three dimensions: \texttt{task}, \texttt{scope}, and \texttt{time\_mode}. These fields are not part of the original benchmark, and we do not treat them as ground-truth annotations. They describe the analytical task expressed by the query, the breadth of the intended view, and the temporality of the analysis. We define them with reference to established visualization taxonomies: low-level analytic task models~\cite{amarLowlevelComponentsAnalytic2005a}, multi-level task and abstraction models~\cite{brehmerMultiLevelTypologyAbstract2013,munzner_nested_2009}, and time-oriented task frameworks~\cite{aignerVisualizationTimeOrientedData2011,schulzDesignSpaceVisualization2013}. This gives us a compact label space for organizing queries in the experiment and guiding retrieval in the grounded setting.
Finally, we construct a stratified evaluation set aimed at queries whose resulting tables are more likely to support multiple reasonable chart encodings. Because each VisEval item resolves to a database result, we base this selection step on the structure of the resulting table. Tables with very low row counts, limited variation, weak categorical structure, or no temporal structure often admit only a narrow range of plausible encodings. We therefore prioritize queries whose tables provide sufficient rows, a clear category structure, and temporal variation to support multiple plausible designs while preserving coverage across the seven chart families represented in VisEval. The resulting evaluation set contains $35$ requests over $31$ databases, with balanced coverage across Bar ($5$), Grouping Line ($5$), Grouping Scatter ($5$), Line ($5$), Pie ($5$), Scatter ($5$), and Stacked Bar ($5$).

\begin{figure}[t!]
\centering
\includegraphics[width=\columnwidth]{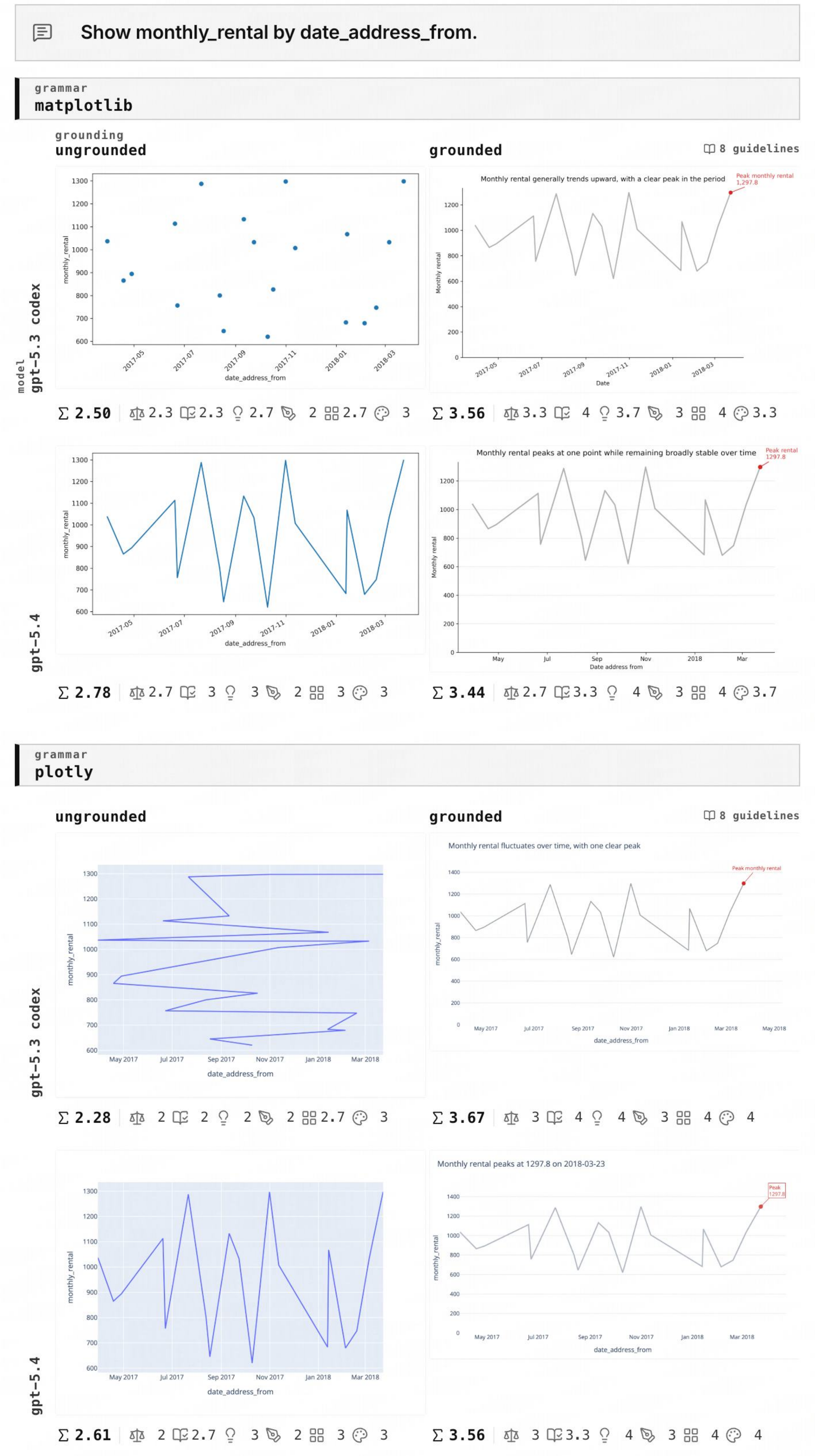}
\vspace{-2em}
\caption[Ungrounded and grounded generation across language models and chart grammars.]%
{Ungrounded and grounded generation for the same analytical query across language models and chart grammars. Groups correspond to grammars, rows to language models, and columns to grounding conditions. The score strip beneath each chart reports average VisJudge scores over three judge runs: \protect\metric{overall}{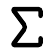}; Faithfulness: \protect\metric{data fidelity}{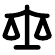}; Expressiveness: \protect\metric{semantic readability}{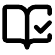}, \protect\metric{insight discovery}{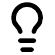}; Aesthetics: \protect\metric{design style}{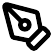}, \protect\metric{visual composition}{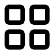}, \protect\metric{color harmony}{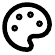}. When the model receives retrieved guidance from the catalog, the resulting designs align more closely across grammars. Without grounding, the same query leads to lower quality scores and more variable chart choices.}
\label{fig:chart-matrix}
\vspace{-3em}
\end{figure}

\vspace{.25em}
\sparagraph{Grounding Conditions}
We compare two grounding conditions. In the ungrounded condition, the model receives only the query and relies on its embedded visualization knowledge. In the grounded condition, the model receives guidance retrieved through our cataloging scheme. This comparison isolates the contribution of our scheme by comparing generation with and without grounding from structured visualization design knowledge.
Grounded retrieval proceeds in two steps. We first use guideline labels to narrow the candidate set based on the query and the resulting table. We then show a retriever LLM (GPT-5.3 Codex) short excerpts for the remaining candidates, built from each guideline's title, description, and \texttt{context} section, and ask it to choose the guidelines that best fit the design problem. This approach prioritizes coverage and, with the tested catalog of $781$ guidelines, each grounding request consumes roughly $160,000$ tokens. Optimizing retrieval for the trade-off between guidance quality and computational cost is outside the scope of this paper. Our goal here is to evaluate the effect of catalog grounding on generative reasoning under conditions where limited retrieval coverage is unlikely to determine the result.
This procedure leverages the catalog's structure. Labels make it possible to narrow the search space, and separate guideline sections make it possible to use different text for different purposes: \texttt{context} helps the model judge whether a guideline applies, while \texttt{advice} and related sections provide the guidance injected into the prompt. The grounded condition therefore supplies a small set of query-specific guidance statements rather than whole documents.

\vspace{.25em}
\sparagraph{Chart Generation and Comparative Evaluation}
For each query-condition pair, we run the same generation loop in both conditions. The model receives the query, a ready-to-visualize table, a textual summary of the table schema and contents, and a list of requirements, then writes Python code in the target grammar. The code is executed, the resulting chart is rasterized, and a vision-language model powered reviewer module checks the code and image against the same requirements and basic visible-quality criteria. If the draft fails, the model receives structured feedback from the reviewer and may revise its code for up to three attempts.
We evaluate only executable outputs. Within generation, drafts are compared using a review-based score that averages five binary checks: requirement compliance, no truncation, no overlap, readable text, and readable marks. This gives the model a limited opportunity to repair obvious implementation problems while keeping the comparison focused on charts that can be rendered and judged.
We use VisJudge-7B from VisJudgeBench~\cite{xieVisJudgeBenchAestheticsQuality2026} for comparative evaluation. Before relying on the released checkpoint, we verify it on the released benchmark charts under a fixed prompt template. This verification asks whether the checkpoint behaves closely enough to the published benchmark to support matched comparative evaluation. The rerun preserves meaningful differences between charts, but it does not reproduce the published benchmark scale exactly. We therefore use VisJudge-7B as a fixed-condition comparative judge, not as absolute ground truth.
For each generated design, we execute the code, rasterize the resulting chart, and submit the image to VisJudge. Each chart is judged three times under the same prompt template. VisJudge evaluates three dimensions of the design: faithfulness, expressiveness, and aesthetics. The six component metrics within these dimensions are listed in the caption of \cref{fig:chart-matrix}. We average each metric over the three runs and then average the six metric scores into a single comparative score, \metric{overall}{overall}, which we use for paired comparisons across conditions and for the later analysis of the consistency of chart-choice across models and grammars.

\vspace{.25em}
\sparagraph{Effects of Grounding on Design Quality and Consistency}
We analyze the effects of grounding as a matched comparison: each grounded chart is compared with the ungrounded chart generated by the same model for a given visualization request, objective, and grammar. The study spans $3$ models, $35$ requests, $2$ objectives (\texttt{refine} and \texttt{select}), $3$ grammars, and $2$ grounding conditions, yielding $1{,}260$ generation attempts. Of these, $1{,}244$ produced executable, rasterizable charts that received VisJudge scores. This yields $613$ complete grounded-versus-ungrounded pairs for analysis, corresponding to $1{,}226$ of the $1{,}260$ total generation attempts. All generated outputs underlying this analysis, together with an interactive viewer for inspecting them, are included in the supplementary material.
Across these matched pairs, grounding increases the overall VisJudge score by $+0.125$ on average (95\% CI [$+0.095$, $+0.155$]). In $434$ pairs ($70.8\%$), the grounded chart scores higher; in $122$ ($19.9\%$), it scores lower; and in $57$ ($9.3\%$), the scores are equal. The effect is positive in both task settings. In \texttt{refine}, grounding increases the score by $+0.132$ (95\% CI [$+0.090$, $+0.174$]). In \texttt{select}, it increases the score by $+0.118$ (95\% CI [$+0.074$, $+0.161$]) and reduces the number of distinct chart types chosen for the same query by $0.343$ on average (95\% CI [$-0.676$, $-0.010$]). The gains are not uniform across the six component metrics of the judge. The largest improvements appear in design style ($+0.441$) and insight discovery ($+0.352$), with semantic readability also improving ($+0.096$), whereas data fidelity shows the weakest effect ($-0.066$). \cref{fig:chart-matrix} illustrates the qualitative side of this result: grounding most clearly improves message clarity, analytical emphasis, and overall design judgment. A detailed trace view of one grounded instance from this matrix appears in \cref{fig:grounded-trace-example}, which shows the retrieved guidelines, their cited sources, and the judge rationales for that chart.

\vspace{.25em}
\sparagraph{Observed Challenges and Improvement Signals}
The empirical study also reveals challenges in grounding generative reasoning. Grounding often constrains the design problem enough to reduce unwarranted variation, but does not always specify the visualization closely enough to determine a single final chart. In many cases, the retrieved guidance fixes the chart family or analytical strategy while leaving finer design choices unresolved, such as vertical versus horizontal layout, grouped versus stacked composition, or the specific takeaway to foreground. \cref{fig:same-guidelines-different-impl-across-models} illustrates this pattern: even when grounded models receive the same guidance and make the same broad structural choice, they can still diverge in how they resolve the remaining ambiguity. This is an inherent trade-off of LLM-driven flows relative to rule-based deterministic systems. The same flexibility that allows a model to adapt retrieved guidance to varied analytical contexts also means that unresolved choices must still be interpreted and instantiated by the model. A second challenge is a mismatch between retrieved guidance and the expressive affordances of the target grammar. As \cref{fig:altair-vs-plotly-non-color-encoding} shows, the same accessibility-oriented guidance can be implemented effectively in Plotly but not as naturally in Altair, leading to substantially different grounded outcomes under otherwise matched conditions. The same traces that reveal weak grounded outcomes also provide a concrete path for improving the catalog. We analyze grounded runs at the level of individual guidelines by measuring how often each retrieved guideline appears and how its presence is associated with the paired grounded-versus-ungrounded score difference, while controlling for objective, model, and grammar. This is possible because the catalog stores guidance as discrete, cited guidelines and each grounded run records which guidelines were retrieved and applied. The resulting diagnostics do not prove causality by themselves, but do localize where improvement effort should go. If weak outcomes cluster in one grammar, the catalog can be revised with grammar-specific applicability notes or grammar-specific variants of the guidance. If a more suitable guideline already exists in the catalog but is not retrieved, the retrieval and ranking procedure should be improved. If the same unmet need recurs and no suitable guideline can be found, the catalog needs additional coverage. If many retrieved guidelines express conceptually identical guidance, they should be consolidated so that grounding remains non-redundant. In this way, weak grounded outcomes do more than show where grounding underperforms: they make the catalog itself inspectable and give a concrete basis for iterative improvement.

\vspace{-0.5em}
\section{Discussion}
\label{sec:discussion}

We position our cataloging scheme as a complement to existing formal systems. The feedback demonstration in \cref{sec:application} illustrates this relationship: VizLinter and Draco returned zero violations for a chart that nonetheless admitted contextually relevant improvements for its specific audience and task. Symbolic constraint systems excel at enforcing structural validity, such as verifying that axes are labeled, encodings are consistent, and visual channels are not overloaded. Our scheme addresses what they cannot: situated guidance that responds to factors such as audience literacy, rhetorical goals, and domain conventions. By layering catalog-grounded retrieval on top of formal validation, systems can combine the precision of symbolic verification with the contextual responsiveness of retrieval-augmented generation.
This separation also addresses a practical concern regarding the grounding of generative models. Providing full papers in LLM contexts has two problems. First, context length is finite; a single paper consumes thousands of tokens. Second, papers optimize for human comprehension. They interleave methodological details, statistical reporting, and discursive framing, diluting the signal for a system seeking actionable design advice.
Academic publishing binds insights to narrative structure---papers shaped by disciplinary conventions and rhetorical forms that serve readers but hinder retrieval. Our cataloging scheme decouples insight from narrative. Treating each guideline as a data object with typed metadata and explicit provenance, we create assets that are independently addressable, semantically retrievable, and usable by both human readers and automated agents. We operationalized this structure in our empirical study (\cref{sec:evaluation}) by retrieving query-relevant guidelines from their \texttt{title}, \texttt{description}, and \texttt{context}, and then injecting the corresponding \texttt{advice} sections to ground generative reasoning. Developing a more efficient retrieval procedure and systematically evaluating retrieval quality for visualization-guideline corpora remain future work.

\vspace{-0.5em}
\subsection{Design Trade-offs and Limitations}

\textbf{Handling Conflicting Guidelines.} The scheme does not enforce consistency. Two guidelines may contradict: one recommends red for emphasis, while the other cautions against it for colorblind accessibility. This permissiveness is intentional. In visualization design, contradictions are typically contextual trade-offs, not logical errors. Forcing resolution at authoring time requires choosing a ``winner'' without knowing the user's situation. Our approach defers this decision to inference time, where the context (audience, task, medium, data type) can break the tie. The operators in \cref{sec:analysis} surface conflicts, presenting competing recommendations to user judgment rather than hiding the tension. In conversational applications, an agent can explain the trade-off and elicit additional context from the designer to guide selection.

\textbf{Traceability under Probabilistic Retrieval.} The scheme relies on embedding-based similarity to match user contexts to relevant guidelines. Embeddings are probabilistic: querying ``elderly readers'' might retrieve guidelines for ``older adults'' or miss relevant entries phrased differently. Does this fuzziness compromise the traceability we claim? Two features mitigate this concern. First, static metadata supports faceted filtering. Deterministic queries on labels like \texttt{audience:older-adults} can narrow the candidate set before semantic search refines it. Second, retrieval returns human-authored text files, not generated content. Retrieval errors surface inapplicable guidelines, not hallucinated ones. Provenance from recommendation to source remains intact; the failure mode is irrelevance, not fabrication.

\textbf{Maintaining the Guideline Catalog.} Automated knowledge extraction creates maintenance burdens. However, explicitly structuring guidelines and labels as semantic data makes this continuous evolution tractable. Recent work utilizing large language models can be leveraged to operationalize this ongoing maintenance. To manage vocabulary drift, models can perform concept induction~\cite{lamConceptInductionAnalyzing2024} to reason over existing structures and propose higher-level abstractions. To handle literature evolution, applying LLM-based ``knowledge delta'' extraction~\cite{el-ebshihyIdentifyingRepresentingKnowledge2023,el-ebshihyBenchmarkCreationNarrative2025} allows the system to systematically track and integrate conceptual differences from new scientific articles. Our scheme provides a substrate for this computational support: algorithms propose candidate merges and knowledge updates, while experts retain the agency to validate them, ensuring the cataloged guidelines remain traceable.

\vspace{-0.5em}
\subsection{Future Work}

The demonstrations in this paper establish feasibility; several directions remain for future exploration.

\textbf{Feedback System Development.} The agentic feedback prototype in \cref{sec:application} demonstrates grounding in a single scenario. A dedicated contribution could develop a full system, investigating prompt engineering, retrieval strategies, and user experience. How should conflicts be surfaced? When should systems seek clarification rather than offer alternatives? What interaction patterns help users understand and trust catalog-grounded recommendations?

\textbf{Process-Level Knowledge.} This paper structures knowledge about visualization decisions: what to do, why it helps, when it applies, and when it breaks down. A complementary layer would represent knowledge about the design process itself. Design study methodologies~\cite{sedlmair_design_2012}, visualization design activity frameworks~\cite{mckenna_design_2014}, and reflective accounts of practitioner workflows~\cite{bigelow_reflections_2014} capture how designers elicit requirements, alternate between divergent exploration and convergent refinement, and iterate with stakeholders. Such process-level knowledge could support agent harnesses that manage multi-step design sessions. Whereas the current catalog guides local design decisions, a process layer would decide which decisions to address next and in what order.

\textbf{Community Infrastructure.} The scheme supports community evolution: natural language lowers authoring barriers, and provenance enables attribution. The same Markdown documents that machines parse render as a browsable website where practitioners explore guidelines by topic, audience, or chart type. Platforms like GitHub could host the catalog where contributors propose additions through pull requests. Continuous integration can automate quality checks: embedding new guidelines at merge time, detecting conflicts with the existing catalog, and flagging duplicates or contradictions. Such infrastructure would transform the catalog from a static artifact to a living resource.

\vspace{-0.5em}
\section{Conclusion}
\label{sec:conclusion}

Automated visualization support navigates a tension between symbolic precision and generative flexibility. Constraint systems enforce structural validity but miss situated nuance; generative models respond to context but lack empirical grounding. We introduced a cataloging scheme that structures design knowledge as natural-language guidelines with typed metadata, i.e., authorable by experts, queryable by machines, and traceable to sources.
18 practitioners confirmed that design reasoning is situated: experts adapt heuristics to audience and intent in ways rigid formalisms cannot capture. We demonstrated expressiveness by cataloging 781 guidelines from perception research, accessibility standards, data journalism, and rhetoric-oriented sources. Embedding sections in a vector space enabled analysis of the represented knowledge itself. Grounding generative reasoning with structured guidelines improved design quality and reduced uncontrolled design variance.
The scheme does not replace existing tools; it provides what they lack. Constraint solvers ensure correctness; the catalog adds context. Generative models produce fluent feedback; the catalog grounds it. Externalizing design knowledge as an inspectable, evolvable resource enables automated support, with recommendations traceable, open to scrutiny, and subject to refinement.

\clearpage

\bibliographystyle{abbrv-doi-hyperref}
\bibliography{template}

\clearpage
\appendix

\begin{figure*}[t]
\centering
\includegraphics[width=\textwidth]{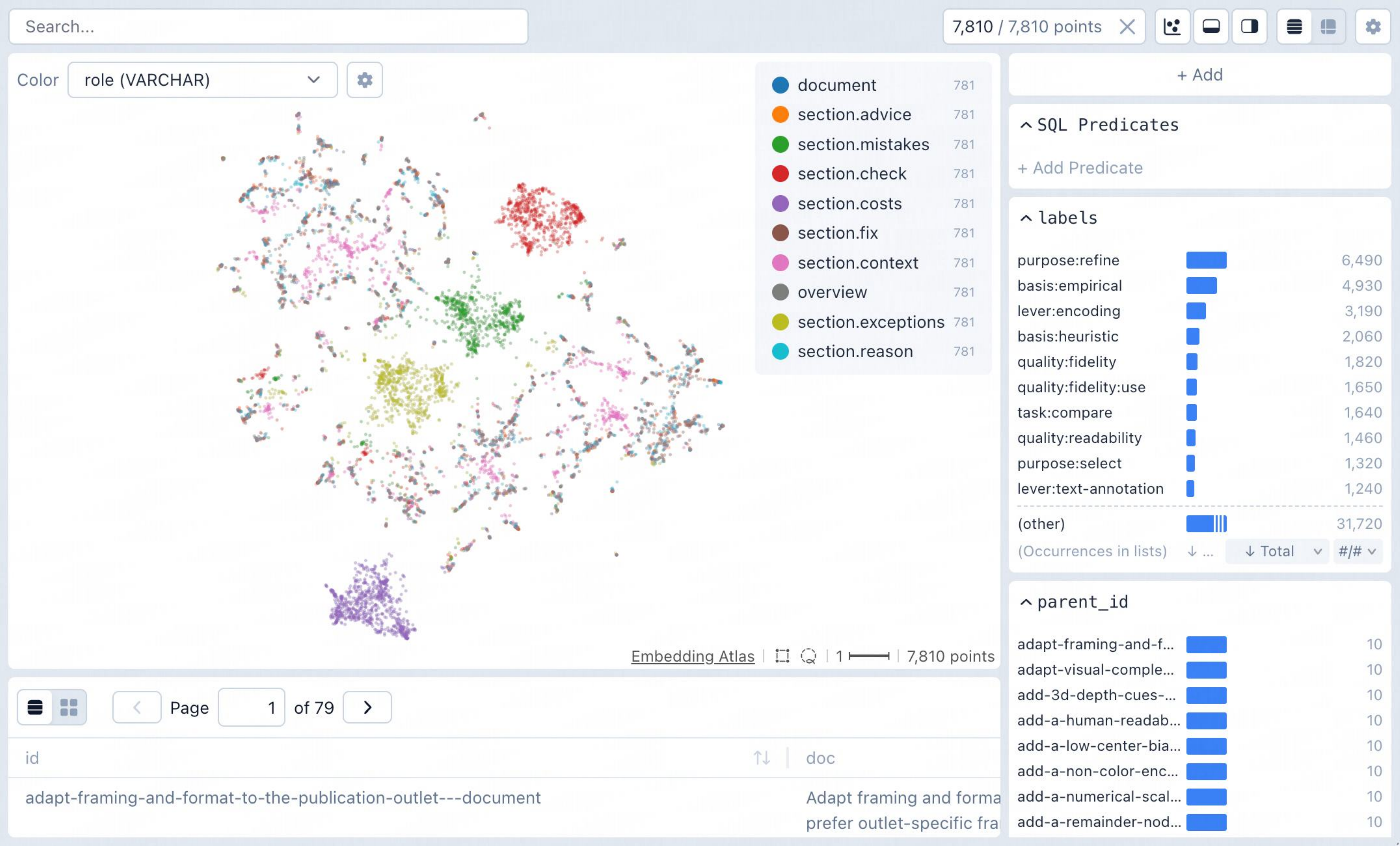}
\caption{Embedding Atlas~~\cite{ren_embedding_2025} view of the $7{,}810$ text embeddings derived from the guideline catalog. The catalog contains $781$ guidelines, and each guideline contributes $10$ embedded texts: the full serialized guideline (\texttt{document}), an overview built from the guideline title and description (\texttt{overview}), and eight role-specific sections (\texttt{section.advice}, \texttt{section.reason}, \texttt{section.context}, \texttt{section.exceptions}, \texttt{section.costs}, \texttt{section.mistakes}, \texttt{section.check}, and \texttt{section.fix}). Points are colored by role. The view makes visible that different representational slices of the same catalog occupy partly distinct regions of the shared embedding space, which motivates our decision to analyze and retrieve sections separately instead of collapsing each guideline into a single monolithic text representation.}
\label{fig:guideline-embedding-atlas}
\end{figure*}

\begin{figure*}[t]
\centering
\includegraphics[width=\textwidth]{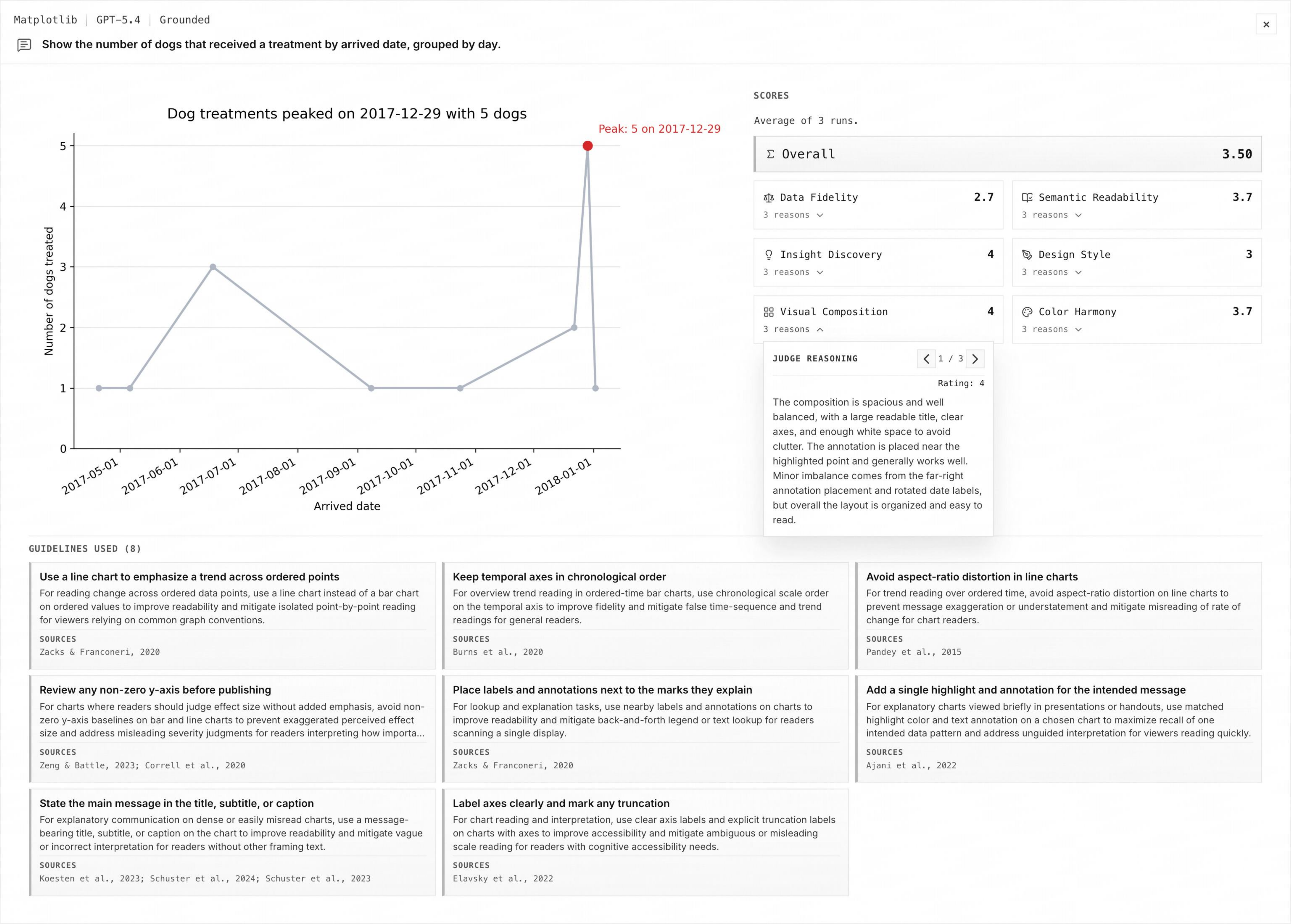}
\caption{Traceability example for a single grounded generation. The figure presents the final chart together with the guideline set retrieved to support it, the source references attached to those guidelines, and the three VisJudge reasoning traces produced across repeated evaluation runs. Taken together, these elements illustrate the full traceability chain enabled by the cataloging scheme: retrieved guidelines remain inspectable as concrete design inputs, their provenance remains visible through cited sources, and the resulting chart can be examined alongside the judge's per-run rationale and aggregated scores.}
\label{fig:grounded-trace-example}
\end{figure*}

\begin{figure*}[t]
\centering
\includegraphics[width=\textwidth]{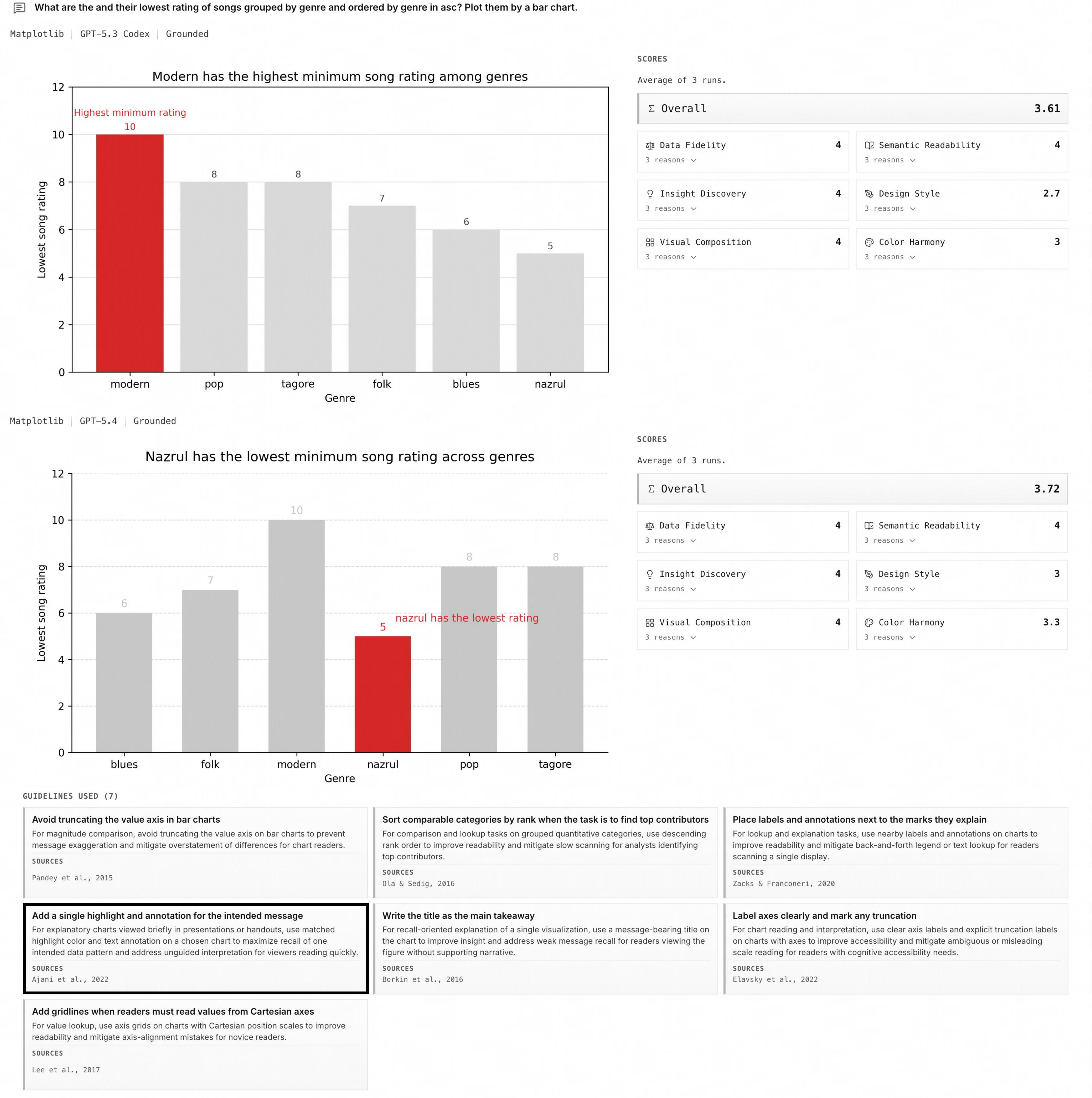}
\caption{Identical analytical intent and retrieved guidance can still yield different grounded implementations across models. Both charts were generated for the same query and conditioned on the same retrieved guidelines, leading to the same broad structural choice while differing in how that guidance is ultimately instantiated. The central ambiguity arises from the highlighting-and-annotation guideline extracted from Ajani et al.~\cite{ajaniDeclutterFocusEmpirically2022}, which recommends adding a single highlight and annotation to foreground the intended message. Because the query does not specify what that message should be, the models resolve the underspecification differently: one treats the highest minimum as the salient takeaway, whereas the other treats the lowest minimum as the key message. This figure thus illustrates a limitation of grounding: the catalog can constrain and document design reasoning, but it does not fully remove residual ambiguity in how a model interprets and applies guidance when the analytical takeaway remains underspecified.}
\label{fig:same-guidelines-different-impl-across-models}
\end{figure*}

\begin{figure*}[t]
\centering
\includegraphics[width=\textwidth]{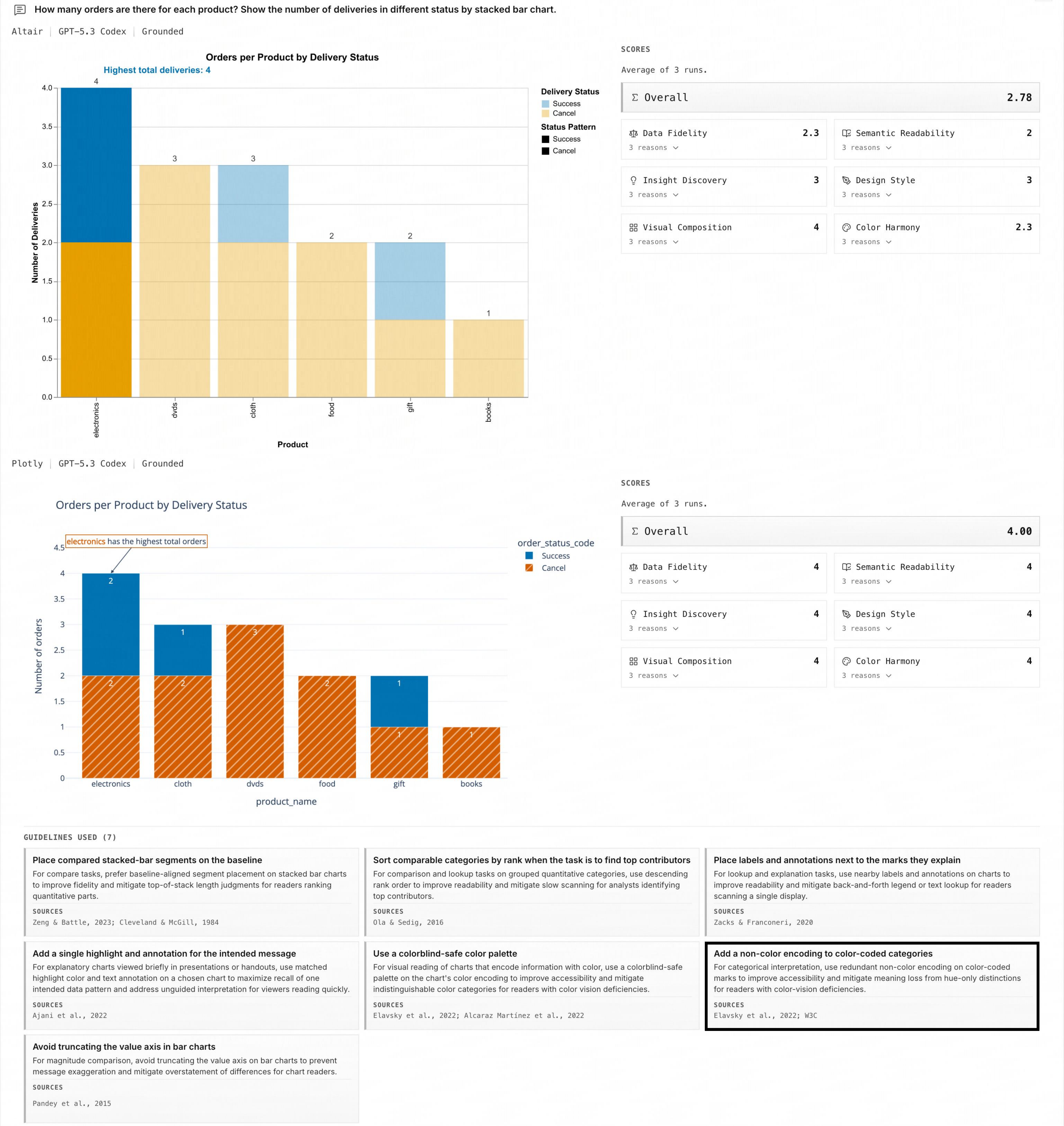}
\caption{Identical analytical query, model, and retrieved guidance can still produce different grounded outcomes when the target grammars afford that guidance differently. In both cases, the retrieved bundle includes an accessibility guideline from Chartability~\cite{elavsky_how_2022} recommending a redundant non-color encoding for color-coded categories so that category membership does not depend on hue alone. In the Plotly chart, the model can satisfy this recommendation with patterned fills, preserving both the stacked-bar structure and the category distinction. In the Altair chart, the model attempts to follow the same guidance, but the available Vega-Lite-based encoding path does not offer an equally direct implementation, leading to a weaker result and a substantially lower VisJudge score. This example shows that some grounded failures arise not from retrieval alone, but from a mismatch between the retrieved guidance and the expressive affordances of the target grammar.}
\label{fig:altair-vs-plotly-non-color-encoding}
\end{figure*}

\end{document}